# Formation of embryos of the Earth-Moon system as a result of a collision of two rarefied condensations

**S. I. Ipatov**[1,2], [1]*Vernadsky Institute of Geochemistry and Analytical Chemistry of Russian Academy of Sciences, Kosygina 19, 119991, Moscow, Russia;* [2]*Space Research Institute of Russian Academy of Sciences, Profsoyuznaya st. 84/32, Moscow, Russia.* Contact: siipatov@hotmail.com

**Main Points of the Abstract:** The angular momentum of the present Earth-Moon system could be acquired at the collision of two identical rarefied condensations with sizes of Hill spheres which total mass was about 0.1 of the mass of the Earth. Solid embryos of the Earth and the Moon could be originated as a result of contraction of the condensation formed at the collision. Depending on eccentricities of planetesimals that collided with solid embryos of the Earth and the Moon, the Moon could acquire 0.04-0.3 of its mass at the stage of accumulation of solid bodies while the mass of the growing Earth increased by a factor of ten.

**Introduction:** Many authors (e.g., [1-3]) suppose that the Earth-Moon system formed as a result of collision of the solid Earth with a Mars-sized object. Galimov and Krivtsov [4] presented arguments that the giant impact concept has several weaknesses. It is considered that one of the weaknesses of the impact theory is that the Moon would consist almost exclusively of material from the impactor, but the rocks from the Moon and the Earth are similar. Ipatov [5] simulated the evolution of a disk of planetesimals initially divided into groups according to their distances from the Sun. He showed that the composition of large enough embryos in the terrestrial feeding zone could be similar due to mixing of planetesimals during planet formation. So in principle, it could be possible that composition of some impactors collided with the embryo of the Earth could not differ much from that of the embryo. Lyra et al. [8] showed that in the vortices launched by the Rossby wave instability in the borders of the dead zone, the solids quickly achieve critical densities and undergo gravitational collapse into protoplanetary embryos in the mass range $0.1M_E$-$0.6M_E$ (where $M_E$ is the mass of the Earth).

Ipatov [6] and Nesvorny et al. [9] supposed that transneptunian satellite systems were formed from rarefied condensations. According to [6], the angular momenta acquired at collisions of condensations moved in circular heliocentric orbits could have the same values as the angular momenta of discovered transneptunian and asteroid binaries. Ipatov [7] obtained that the angular momenta used in [9] as initial data in calculations of the contraction of condensations leading to formation of transneptunian binaries could be acquired at collisions of two condensations moved in circular heliocentric orbits. Ipatov supposed that the number of collisions of condensations at which the formed condensation with mass equal to that of a solid body with diameter $d$>100 km got the angular momentum needed for formation of a satellite system can be about the number of small bodies with $d$>100 km having satellites, i.e., the fraction of condensations formed at collisions leading to formation of satellite systems among all condensations can be about 0.3 for solid primaries with $d$>100 km formed in the transneptunian belt. The model of collisions of condensations explains negative angular momenta of some observed binaries (e.g., 2000 CF105, 2001 QW322, 2000 QL251), as about 20 percent of collisions of condensations moving in circular heliocentric orbits lead to retrograde rotation.

**The Angular Momentum at a Collision of Two Condensations**: Using the formulas presented in [6], we obtained that the angular momentum $K_{EM}$ of the Earth-Moon system equals to the angular momentum $K_{s2}$ at a typical collision of two identical condensations with size of Hill spheres, which total mass equals $0.13M_E$. For circular heliocentric orbits, the maximum value of $K_{s2}$ is greater by a factor of $0.6^{-1}$ than the above typical value [6]. In this case, the above total mass is $0.096M_E$. Therefore, the angular momentum of the Earth-Moon system could be acquired at a collision of two condensations with a total mass not smaller than $0.1M_E$. We suppose that solid proto-Earth and proto-Moon could form by contraction of a condensation (e.g., according to the models of contraction of a condensation presented in [4, 9]). Not all material of collided preplanetesimals could be left in the condensation formed at the collision. So the total mass of collided preplanetesimals could exceed $0.1M_E$.

**Angular Velocities of Condensations Needed for Formation of Satellite Systems:** In calculations of contraction of condensations (of mass $m$ and radius $r$ equal to 0.6 of the Hill radius $r_H$) presented in [9], trans-Neptunian objects with satellites were formed at initial angular velocities $\omega_o$ from the range $0.5\Omega_o$–$0.75\Omega_o$, where $\Omega_o=(Gm/r^3)^{1/2}$ ($G$ is the gravitational constant). In 3-D calculations of gravitational collapse of a condensation presented in [4], binaries were formed at $\omega_o/\Omega_o$ from the range of 1-1.46. For smaller $\omega_o/\Omega_o$, satellites were not formed, and a considerable fraction of angular momentum could be in particles that left the formed condensation. The difference in results presented in [4] and [9] can be caused, in particular, by different chaotic velocities of particles/bodies constituting condensations and by different sizes of condensations. The sizes of condensations in calculations presented in [4] were smaller than Hill spheres. For example, in one of their calculations (page 108 in [4]) the radius of the condensation exceeded the radius of the corresponding solid planet by a factor of 5.5, while the Earth radius is smaller by a factor more than 200 than the Hill radius.

At a collision of two identical condensations, the angular velocity can be as high as $\omega=1.575\Omega$, with a mean value of about $\omega\approx0.945\Omega$, where $\Omega=(G\cdot M_S)^{1/2}a^{-3/2}$ is the angular velocity of a condensation around the Sun [6]. As $\Omega_o/\Omega\approx1.73(r_H/r)^{3/2}$, then $\Omega_o\approx1.73\Omega$ at $r=r_H$. In this case $\omega$ can be as high as $0.9\Omega_o$, but it is smaller for smaller $r$.

Collided condensations could have non-zero angular momenta before the collision. So in principle the resulting angular velocity of the formed condensation can exceed $0.9\Omega_o$. According to Safronov [10], the initial angular velocity $\omega_o$ of a rarefied condensation is $0.2\Omega$ for a sperical condensation and $0.25\Omega$ for a flat circle. The initial angular velocity is positive and is not enough for formation of satellites. If two identical uniform spherical condensations with $\omega_o$ collided without additional angular momentum, then the angular velocity of the spherical condensation formed at the collision is $\omega_2=2^{-2/3}\omega_o$; e.g., $\omega_2=0.126\Omega$ at $\omega_o=0.2\Omega$. The ratio of $\omega/\omega_2=0.945/0.126=7.5$ shows that in the case of a collision of equal mass condensations the increment to the angular velocity due to a typical collision is greater by a factor of several than the angular velocity which is due to initial rotation of condensations. The angular velocity of a condensation of radius $r_c$ formed as a result of compression of the condensation, with the Hill radius $r_H$ and the angular velocity $\omega_H$, equals $\omega_{rc}=\omega_H(r_H/r_c)^2$ [6]. Therefore any initial angular velocities considered in [4,9] can be reached after contraction of the condensation formed at a collision of condensations not greater than Hill spheres.

**The Growth of Solid Embryos of the Earth and the Moon:** Galimov and Krivtsov [4] studied the growth of the rotating planet-satellite system in the accumulation of the matter of the dust cloud. In particular, they obtained that the present masses of the Earth-Moon system can be got at the initial total mass of the system equal to 0.047 of the present value and at the initial ratio of the embryos equal to 4.07 (for the growth of masses of the Earth and the Moon by a factor of 26.2 and 1.31, respectively). These studies were made for the model when velocities of incoming dust were zero at the edges of a cylinder around the Earth-Moon system. In the case of eccentrical heliocentric orbits of particles, this model is not true. In simulations presented in [5], mean eccentricities of planetesimals in the terrestrial zone exceeded 0.2 (and later 0.3) at some stages of evolution.

Let us consider the model of the growth of solid embryos of the Earth and the Moon to the present masses of the Earth and the Moon ($M_E$ and $0.0123M_E$, respectively) by accumulation of smaller planetesimals for the case when the effective radii of proto-Earth and proto-Moon are proportional to $r$ (where $r$ is the radius of a considered embryo). Such proportionality can be considered for large enough eccentricities of planetesimals. In this case, based on $dm_M/m_M=k\cdot(m_M/m_E)^{2/3}dm_E/m_E$ we can obtain $r_{Mo}=m_{Mo}/M_E=[(0.0123^{-2/3}-k+k\cdot(m_{Eo}/M_E)^{-2/3})]^{-3/2}$, where $k=k_d^{-2/3}$, $k_d$ is the ratio of the density of the growing Moon of mass $m_M$ to that of the growing Earth of mass $m_E$ ($k_d=0.6$ for the present Earth and Moon), $m_{Mo}$ and $m_{Eo}$ are initial values of $m_M$ and $m_E$. For $r_{Eo}=m_{Eo}/M_E=0.1$, we have $r_{Mo}=0.0094$ at $k=1$ and $r_{Mo}=0.0086$ at $k=0.6^{-2/3}$. At these values of $r_{Mo}$, the ratio $f_m=(0.0123-r_{Mo})/0.0123$ of the total mass of planetesimals that were accreted by the Moon at the stage of the solid body accumulation to the present mass of the Moon is 0.24 and 0.30, respectively. In this case for the growth of the mass of the Earth embryo by a factor of ten, the mass of the Moon embryo increased by a factor of 1.31 and 1.43, respectively.

If we consider that effective radii of the embryos are proportional to $r^2$ (the case of small relative velocities of planetesimals), then integrating $dm_M/m_M=k_2\cdot(m_M/m_E)^{4/3}dm_E/m_E$, we can get $r_{Mo2}=m_{Mo}/M_E=[(0.0123^{-4/3}-k_2+k_2\cdot(m_{Eo}/M_E)^{-4/3})]^{-3/4}$, where $k_2=k_d^{-1/3}$. In the case of $r_{Eo}=m_{Eo}/M_E=0.1$, we have $r_{Mo}=0.01178$ at $k_2=1$ and $r_{Mo}=0.01170$ at $k_2=0.6^{-1/3}$, and $f_m$ equals 0.042 and 0.049, respectively. In this case for the growth of the Earth embryo mass by 10 times, the Moon embryo mass increased by the factor of 1.044 and 1.051 at $k_2=1$ and $k_2=0.6^{-1/3}$, respectively. In the above model, depending on eccentricities of planetesimals, the Moon could acquire 0.04-0.3 (the lower estimate is for almost circular heliocentric orbits) of its mass at the stage of accumulation of solid bodies during the time when the mass of the growing Earth increased by a factor of ten. Probably the initial mass of a solid proto-Earth could exceed $0.1M_E$, and so the growth of the Moon embryo could be smaller than the estimate obtained for the growth of the mass of the Earth embryo by a factor of ten.

We suppose that the condensations that contracted and formed the embryos of other (not the Earth) terrestrial planets did not collide with massive condensations, and therefore the condensations formed at collisions did not get large enough angular momentum needed to form massive satellites.

This study was supported by Program no. 22 of the Presidium of the Russian Academy of Sciences.